\documentclass[3p,times,procedia]{elsarticle}
\usepackage{ecrc}
\usepackage{caption}
\usepackage[bookmarks=false]{hyperref}
    \hypersetup{colorlinks,
      linkcolor=blue,
      citecolor=blue,
      urlcolor=blue}
\usepackage{amsmath}

\volume{00}

\firstpage{1}

\journalname{Procedia Computer Science}

\runauth{Liu, Zheng, and Cartlidge}

\jid{procs}

\usepackage{amssymb}
\usepackage[figuresright]{rotating}
\usepackage{booktabs} % for tables

\begin{document}
\begin{frontmatter}

%% Title, authors and addresses

%% use the tnoteref command within \title for footnotes;
%% use the tnotetext command for the associated footnote;
%% use the fnref command within \author or \address for footnotes;
%% use the fntext command for the associated footnote;
%% use the corref command within \author for corresponding author footnotes;
%% use the cortext command for the associated footnote;
%% use the ead command for the email address,
%% and the form \ead[url] for the home page:
%%
%% \title{Title\tnoteref{label1}}
%% \tnotetext[label1]{}
%% \author{Name\corref{cor1}\fnref{label2}}
%% \ead{email address}
%% \ead[url]{home page}
%% \fntext[label2]{}
%% \cortext[cor1]{}
%% \address{Address\fnref{label3}}
%% \fntext[label3]{}

\dochead{24th International Conference on Modelling and Applied Simulation (MAS 2025), held within the 22nd International Multidisciplinary Modeling \& Simulation Multiconference (I3M 2025)}%%%
%% Use \dochead if there is an article header, e.g. \dochead{Short communication}
%% \dochead can also be used to include a conference title, if directed by the editors
%% e.g. \dochead{17th International Conference on Dynamical Processes in Excited States of Solids}

\title{Deep Reinforcement Learning for Optimal Asset Allocation Using DDPG with TiDE}

%% use optional labels to link authors explicitly to addresses:
%% \author[label1,label2]{<author name>}
%% \address[label1]{<address>}
%% \address[label2]{<address>}

% %%% Use the \authfn to add symbols for additional footnotes, if any. 1 is reserved for correspondence emails.
% \author[1,\authfn{1}]{Rongwei Liu}
% \author[2]{Jin Zheng}
% \author[2]{John Cartlidge}

% \affil[1]{School of Computer Science, University of Bristol, UK}
% \affil[2]{School of Engineering Mathematics and Technology, University of Bristol, UK}

% %%% Corresponding author
% \authnote{\authfn{1}Corresponding author. Email address: rongwei.liu@bristol.ac.uk}

\author[a]{Rongwei Liu\corref{cor1}} 
\author[b]{Jin Zheng}
\author[b]{John Cartlidge}

\address[a]{School of Computer Science, University of Bristol, Bristol, UK}
\address[b]{School of Engineering Mathematics and Technology, University of Bristol, Bristol, UK}

\begin{abstract}
The optimal asset allocation between risky and risk-free assets is a persistent challenge due to the inherent volatility in financial markets. Conventional methods rely on strict distributional assumptions or non-additive reward ratios, which limit their robustness and applicability to investment goals. To overcome these constraints, this study formulates the optimal two-asset allocation problem as a sequential decision-making task within a Markov Decision Process (MDP). This framework enables the application of reinforcement learning (RL) mechanisms to develop dynamic policies based on simulated financial scenarios, regardless of prerequisites. We use the Kelly criterion to balance immediate reward signals against long-term investment objectives, and we take the novel step of integrating the Time-series Dense Encoder (TiDE) into the Deep Deterministic Policy Gradient (DDPG) RL framework for continuous decision-making. We compare DDPG-TiDE with a simple discrete-action Q-learning RL framework and a passive buy-and-hold investment strategy. Empirical results show that DDPG-TiDE outperforms Q-learning and generates higher risk adjusted returns than buy-and-hold. These findings suggest that tackling the optimal asset allocation problem by integrating TiDE within a DDPG reinforcement learning framework is a fruitful avenue for further exploration.
\end{abstract}

%% Type your keywords here, separated by semicolons ; 
\begin{keyword}
Optimal Asset Allocation; Kelly Strategy Simulation; Reinforcement Learning; Q-Learning; DDPG; TiDE; Sharpe Ratio

%% keywords here, in the form: keyword \sep keyword

%% PACS codes here, in the form: \PACS code \sep code

%% MSC codes here, in the form: \MSC code \sep code
%% or \MSC[2008] code \sep code (2000 is the default)
\end{keyword}

\cortext[cor1]{Corresponding author.}
\end{frontmatter}

%\correspondingauthor[*]{Corresponding author. Tel.: +0-000-000-0000 ; fax: +0-000-000-0000.}
\email{rongwei.liu@bristol.ac.uk}

%%
%% Start line numbering here if you want
%%
% \linenumbers
\vspace*{-6pt}
%% main text

%\enlargethispage{-7mm}
\section{Introduction}
\label{sec:introduction}
\noindent
Financial markets comprise a diverse array of assets competing for investors' limited capital. Asset returns, characterized by volatility, do not necessarily equate to a market risk premium but rather reward exposure to systematic risk. Diversified portfolios aid in managing idiosyncratic risk, with the market portfolio embodying non-diversifiable uncertainty that influences asset prices. While the true market portfolio remains unobservable, it is commonly approximated by a broad stock market index. This study centers on optimizing asset allocation between a risk-free asset and the market portfolio. Rather than estimating the efficient frontier or determining the exact tangency point of the Capital Market Line, the focus is on identifying the investor-specific optimal allocation to maximize long-term portfolio performance, assuming the market portfolio is observable and broadly representative of the investable universe.

The classical optimal portfolio methods are mean variance optimization (MVO) \citep{Markowitz1952}, which was proposed to construct portfolios. Several tools have been used to estimate the distribution of returns on assets and to select valuable portfolios. Because the true distribution of returns is unknown, the population is estimated and substituted by the sample, and then the optimal portfolio weights are selected based on the conclusions derived. The solution of portfolio weights that maximize utility depends on the first-order and second-order returns moments and the investor's risk aversion coefficient \citep{KanZhou2007Optimal}. This method is known as plug-in estimation. The limitation of this approach is that it assumes a consistent estimate of population statistics, but the estimation results will not converge when the assets in the portfolio gradually increase. 

The hyperparameters of machine learning (ML) models can be tuned to maximize the expected out-of-sample utility or minimize a predefined risk function, provided that the first-order and second-order moments and the functional form of the return distribution are known \citep{Pinelis2022Machine}. First, a class of equations with hyperparameters is selected, the samples are trained, and the hyperparameters are estimated. Second, the hyperparameters are tuned with out-of-sample data on the validation set to get the best performance. This is a two-step approach. For portfolio selection methods, cutting-edge machine learning directly considers the integration of utility maximization into statistical problems for weighted function estimation \cite{Zhang2021UniversalE2E}. This method is known as the one-step procedure or parametric portfolio weight approach. It is not necessary to know the distribution assumptions of the returns and, instead, only the investment utility function and the weight function are needed. An intuitive example is the Maximum Sharpe Ratio Regression (MSRR) \citep{Britten2002MV}, which is adaptable to sophisticated machine learning models such as neural networks \citep{Simon2022DeepPP}. 

Reinforcement learning (RL) uses a reward signal from a simulated market environment to iteratively guide learning, and has been widely used in many problems, including asset allocations \cite{Hambly2023Recent}. In this paper, we introduce the utility equation based on the Kelly criterion and determine the reward through formula derivation to establish the optimal policy problem within the MDP framework. Additionally, we take the novel approach of incorporating the Time-series Dense Encoder (TiDE) \cite{Das2023LongtermFW} into the Deep Deterministic Policy Gradient (DDPG) \cite{Lillicrap2015ContinuousCW} mechanism to enhance the handling of temporal relationships in multivariate time series within the state space. 

The remainder of this paper is organized as follows. Section~\ref{sec:State of the art} reviews the classification of RL, the research gaps in actor-critic approaches, and the theoretical basis and methodological innovations in the solutions proposed in this study. Section~\ref{sec:Methods and Materials} elucidates the optimal function according to the Kelly criterion within the MDP framework and explains the iterative principle behind DDPG with TiDE. The source data, experimental configurations, and performance metrics are also described. Section~\ref{sec:results} presents the experimental results. Finally, Section~\ref{sec:conclusion} concludes and suggests potential avenues for future work.

\section{State of the art}
\label{sec:State of the art}
\noindent
RL encompasses various algorithms distinguished by interaction paradigms (online vs. offline), policy update mechanisms (on-policy vs. off-policy), and model reliance (model-based vs. model-free). Although this classification offers a useful conceptual framework, its practical implications are particularly significant in financial contexts. Online RL, though theoretically appealing for its adaptability, can encounter latency and instability due to the high costs and risks associated with live market interactions. Conversely, offline RL allows learning from historical or simulated data, but struggles with generalization if the training data are not adequately representative. On-policy methods learn from the policy they aim to improve, ensuring internal consistency but limiting exploration. In contrast, off-policy methods like Q-learning \cite{Watkins1992Q} use data from different behavior policies, enhancing data efficiency, a crucial advantage in finance where real-time sampling is limited \citep{Park2020AIFP} \citep{Pendharkar2018Trading}. Therefore, this study adapts the ideas underlying the success of Q-learning to the off-policy domain as the starting point for simulation. Model-based RL approaches, compared to model-free mechanisms, remain largely confined to simulating and modeling price–volume data and continue to suffer from the inherent suboptimality of the two-step process of prediction followed by optimization. Due to variations in data structures and forecasting models, the robustness and generalizability of such methods remain subject to further scrutiny \citep{Wei2019ModelbasedRL} \citep{Yu2019ModelbasedDR}. The datasets used in this paper cover more than twelve indicators. Consequently, this study focuses on model-free RL, which prevails in financial contexts. Within this domain, algorithms are generally divided into actor-only and critic-only approaches according to how the architectures represent decision-making. Actor-only methods (e.g., policy gradient) directly optimize policies but suffer from high variance and slow convergence, especially in noisy settings. In contrast, critic-only methods (e.g., Q-learning) offer greater stability through value estimation, but struggle with flexibility in continuous action spaces. Despite extensive study, the comparative advantages of these paradigms remain unclear due to inconsistent benchmarks and experimental setups in the literature \cite{Bhakar2023MaxRet} \cite{Zhao2025QuantFactor}.

Actor-Critic (AC) architectures integrate policy learning (actor) with value estimation (critic) to improve sample efficiency and facilitate continuous decision-making \citep{Lillicrap2015ContinuousCW}. A notable implementation within this framework is the DDPG mechanism, which employs neural networks for both actor and critic roles. The application of DDPG to portfolio optimization has been widely studied. Jiang et al. \cite{Jiang2017ADR} introduced an RL framework incorporating CNN, RNN, and LSTM networks, all of which outperformed traditional benchmarks in cryptocurrency markets. Liang et al. \cite{Liang2018AdversarialDR} compared DDPG, PPO, and PG on five randomly selected stocks from the Chinese market, concluding that PG was preferable. Pham et al. \cite{Pham2021MultiagentRL} investigated multi-agent RL using 10 popular stocks from the Ho Chi Minh Stock Exchange and VN30 Index Futures, demonstrating superior performance over the risk-free rate and the potential for cross-hedging strategies. Cong et al. \cite{Cong2021Alphaportfolio} developed AlphaPortfolio, a multi-sequence, attention-based model for direct investment target optimization, finding that the Sharpe ratio is time delayed compared to other reward settings. Aboussalah et al. \cite{Aboussalah2021What} applied CNN-based RL to asset allocation, noting limited success with stable algorithms and proposing a more interpretable strategy by integrating the Kelly criterion. Although DDPG has been widely applied in asset allocation tasks, many studies are limited to specific markets or rely on a small subset of randomly chosen stocks. This narrow focus undermines the generalizability of their conclusions and limits their relevance to broader investor populations. Moreover, a critical shortcoming in the existing application of RL to portfolio optimization lies in the insufficient attention paid to reward design within the MDP framework. Many prior works implicitly assumed that maximizing cumulative Sharpe ratios is equivalent to identifying an optimal strategy but have failed to account for the non-additive and time-lagged nature of this metric, potentially leading to mismatches between the learning signal and true investor objectives. 

This study seeks to fill the above research gaps by examining portfolio allocation between a market index---the CRSP index, serving as a representative proxy for the aggregate U.S. equity market---and a risk-free asset. To ensure theoretical rigor, the study adopts the Kelly criterion through the use of the constant relative risk aversion (CRRA) utility function and the accumulated reward, providing mathematical consistency between log-utility maximization and long-term capital growth. Furthermore, to better capture temporal dependencies among heterogeneous financial indicators, we take the novel step of integrating the TiDE model into the DDPG framework.\footnote{On 25/06/2025, R. Liu conducted a review of all 472 papers listed on GoogleScholar as papers that cited the original TiDE article \cite{Das2023LongtermFW}. A total of 440 articles were accessible to read: of these, 240 articles implemented TiDE in some capacity, while the others only mentioned TiDE in discussion. Of the studies that implemented TiDE, none incorporated TiDE within an RL framework and only 12 papers mentioned RL. Therefore, we conclude that no previous publication has used TiDE within the DDPG framework for RL.} This architectural enhancement enables the agent to process richer financial information and learn more adaptive and robust portfolio policies. By conducting comparative experiments using DDPG with TiDE, Q-learning, and a buy-and-hold benchmark, we will assess the capacity of RL to generate investor-specific optimal allocations under long-horizon utility-based criteria. This establishes a stronger foundation for the development of data-driven, adaptive portfolio strategies in realistic financial environments.

\section{Methods and Materials}\label{sec:Methods and Materials}

\subsection{Kelly Strategy and MDP Simulation}
\noindent
In 1971, Merton \cite{Merton1971Optimum} explored the trading of stocks and bonds within his portfolio model to optimize the utility of investors. In 2010, MacLean \cite{Maclean2010Medium} equated the investor's optimal expected utility with the market's maximum expected return, labeling the portfolio's investment objective as the Kelly criterion, formulated as $\max E[\ln U(W)] = \max E[\ln Return]$, where $Return=1+r_t$ and $r_t$ is the rate of risky asset return $\sim\mathcal{N}\left(\mu, \sigma^{2}\right)$. Then, based on the derivation using It\^{o}'s Lemma, the optimal investment fraction in the risky asset is calculated as $\pi^{*}=(\mu-r_f)/R_a \sigma^{2}$, where $\mu$ is the expected return of the risky asset, $r_f$ is the return rate of the risk-free asset, $\sigma$ is the standard deviation, $R_a$ is the coefficient of constant risk aversion. However, the distribution of risky assets in real financial markets is complex and unknown, and its indicators are not limited to price data. Therefore, the optimal solution is limited by various conditions and cannot adapt to the dynamically adjusting market environment. Using MDP, RL algorithms are developed for the practical implementation of the RL Kelly strategy \cite{Jiang2022Kelly}.

An MDP provides a mathematical framework for sequential decision making under uncertainty, defined by the tuple (S, A, P, R, $\gamma$), where S is the state space, A is the action space, P(s' $\mid$ s, a) denotes transition probabilities, R(s, a) is the reward function, and $\gamma$ is the discount factor. In RL, an agent interacts with an MDP environment, learning from rewards and iteratively improving its policy. Within financial markets, the environment can be treated as an MDP whose states capture market conditions (e.g., historical macro factors, excess returns, and volatilities), whose actions represent trading decisions (e.g., buy, sell, hold, or adjusting risk-asset weights), and whose rewards measure long-term value (e.g., log-return). By simulating a variety of historical or synthetic market scenarios, an RL agent explores possible strategies, continuously refines its portfolio allocations through reward feedback, and strives to maximize cumulative logarithmic returns while respecting risk constraints. Through repeated trials, RL avoids strict rule-based heuristics and adapts dynamically to changing market conditions.

\subsection{CRRA Utility and Reward Setting}
\noindent
In RL for portfolio optimization, the reward must provide immediate feedback on the chosen action but also embody the longer-term risk-return trade-off. One concise way to encode both dimensions is through a coefficient of constant relative risk-aversion (CRRA)\footnote{CRRA is a widely used assumption in economic models; and is assumed in Merton's model \cite{Merton1969Lifetime}.} utility function:
\begin{equation}
     U(W)=\frac{W^{1-R_a}}{1-R_a}
\end{equation}
where $W$ is wealth and $R_a$ is relative risk aversion (constant). When this expression reduces to logarithmic utility, then:
\begin{equation}
    \ln U(W)=(1-R_a)\ln (W)-\ln (1-R_a)
\end{equation}
Maximizing the expectations of both sides, since $0<R_a<1$, yields the following:
\begin{equation}
    \max E[\ln U(W)]=\max E[(1-R_a)\ln (W)-\ln (1-R_a)]=\max E[\ln (W)] 
\end{equation}

\noindent Hence, setting, 
\begin{equation}
    Reward_t=\gamma \ln (1+r_{t})
\end{equation}
gives total reward,
\begin{equation}
    \sum_{t=1}^{T}  Reward_t = \sum_{t=1}^{T} \ln (1+r_t)^{\gamma} = \ln \prod_{t=1}^{T} (1+r_t)^{\gamma}
\end{equation}
and therefore wealth at time T is,  
\begin{equation}
    W_0 \prod_{t=1}^{T} (1+r_t)^{\gamma} = W_T ,  \text { where assuming } W_0=1.
\end{equation}

Hence, to maximize expected wealth, we have:
\begin{equation}
    \max E[\ln W_0+\sum_{t=1}^{T}  Reward_t ]= \max E[ \ln W_0\prod_{t=1}^{T} (1+r_t)^{\gamma}]=\max E[\ln W_t]
\end{equation}
which corresponds exactly to a log-utility perspective $\max E[\ln U(W)]$. Optimizing the expected sum (or discounted sum) of rewards aligns with maximizing the expected future log wealth, known as the Kelly criterion \citep{Maclean2011Longterm}. This formulation ensures that each action's return-risk balance (via the immediate log-return feedback) contributes to the long-run objective of sustainable capital growth, thereby unifying immediate step-by-step reward signals with a principled, long-horizon utility measure.

\subsection{Q-Learning and Bellman Equation}
\noindent
Since the true transition dynamics of time series is typically unknown, Q-learning \cite{Watkins1992Q} uses temporal difference (TD) learning to iteratively approximate $Q^*(s, a)$ using sampled experience $(s, a, r, s')$. The update rule follows from the Bellman equation:

\begin{equation}
    Q^*(s, a) = \mathbb{E} \left[ r + \gamma \max_{a'} Q^*(s', a') \mid s, a \right]
\end{equation}

The Bellman equation is the foundation of Q-learning, providing a recursive relationship to estimate the optimal action-value function $Q^*(s, a)$. The Bellman equation expresses the value of a state-action pair as the immediate reward plus the expected discounted return of the optimal future actions:

\begin{equation}
    Q(s, a) \leftarrow Q(s, a) + \alpha \left( r + \gamma \max_{a'} Q(s', a') - Q(s, a) \right)
\end{equation}
\noindent
where $\alpha$ is the learning rate that controls the magnitude of the update; $\gamma$ is the discount factor that weighs future rewards; and $r + \gamma \max_{a'} Q(s', a')$ is the TD target, representing the optimal future return estimate. This update process ensures that Q-learning converges to $Q^*(s, a)$ with adequate exploration and appropriate scheduling of learning rates. In practice, Q-learning is particularly useful for discrete action spaces, such as financial trading decisions (buy/sell/hold), where it helps optimize trading strategies by learning from simulated market interactions.

In this work, the Q-learning algorithm is applied with a discretized observation and action space to balance the tractability of the model with sufficient flexibility. Specifically, the state space is obtained by performing K-means clustering ($k=50$) on key market features, yielding a finite set of cluster labels as the observed states. For actions, rather than using a simplistic buy-sell-hold triad, the agent discretizes portfolio weights $\omega \in \{0.0, 0.1,\ldots,1.0\}$. This design allows for somewhat finer control over investment levels, while remaining computationally manageable. As a benchmark, this study includes the buy-and-hold strategy---invest all funds in the market index at the beginning and hold until the end---as a passive baseline with fixed $\omega=1.0$. By comparing against a passive strategy, we can ensure that any gains in performance achieved by Q-learning are due to actively rebalancing portfolio weights.

\subsection{DDPG with TiDE}
\noindent
Q-learning is a value-based RL algorithm designed for discrete state-action spaces. In the described experiments, the market is represented by a K-means-based discretization into 50 clusters and logarithmic utility (Kelly criterion) is optimized. However, this approach suffers from two major challenges: the Q-table becomes impractically large as the number of clusters increases, and action selection by maximizing Q-values entails a costly search over all discrete bins. To address these limitations, policy-based methods were introduced, allowing RL agents to learn a policy $\pi(a|s)$ directly rather than relying on Q-values. The DDPG algorithm extends these ideas by combining the value function learning of Q-learning with policy optimization, making it suitable for continuous action spaces. Unlike stochastic policy-based methods, DDPG learns a {\em deterministic policy} function $a = \pi_{\theta}(s)$ that maps each multivariate time series state directly to a continuous action, thus avoiding an expensive search for action selection.

Within the DDPG {\em actor-critic} framework, an actor network $\pi_{\theta}(s)$ outputs portfolio weights and a critic network $Q_{\phi}(s, a)$ evaluates them. The actor is then updated using the deterministic policy gradient:

\begin{equation}
    \nabla_{\theta} J = \mathbb{E} \left[ \nabla_a Q_{\phi}(s, a) \nabla_{\theta} \pi_{\theta}(s) \right]
\end{equation}

\noindent
which ensures that selected actions maximize the critic’s Q-value. The critic is updated using Bellman's equation, employing a soft‐updated target critic $Q_{\phi}(\tau = 0.005)$ to enhance stability. 

The agent adopts a TiDE-based architecture integrated within the DDPG framework. Historical features are passed to a two-layer TiDE encoder, which flattens the inputs and projects them into a latent representation using fully connected layers followed by residual blocks with ReLU activation and layer normalization. The TiDE actor maps the encoded state into a scalar action through a final linear transformation and a sigmoid activation, producing portfolio weights $\omega \in[0,1.0]$. The TiDE critic evaluates the quality of the state-action pair by concatenating the TiDE encoder's output with the TiDE actor's action and processes it through two fully connected layers with ReLU activations to output the Q-value. This encoder-based actor-critic structure enables the model to handle multivariate time series inputs efficiently, while the deterministic policy of DDPG facilitates continuous portfolio weight selection, offering finer control and reduced computation compared to discrete action methods.

\subsection{Dataset and Experiments}
\noindent
We used a dataset consisting of monthly time series on the market return rate $r_t$ (Mkt) and risk-free asset return rate $r_f$ (RF) from 1927 to 2019. We separated into an initial training period (1927--1957), a validation period (1958--1988), and a test period (1989--2019). Daily returns are retrieved to compute the monthly realized volatilities. The original data set library is available on the Fama/French website.\footnote{Data in Fama/French 3 Factors csv, see: \url{https://mba.tuck.dartmouth.edu/pages/faculty/ken.french/data_library.html}.}
There are 11 macroeconomic and financial predictors for market environment indicators, shown in Table~\ref{tab:predictors}. An additional variable is the one-month lagged excess return. Lastly, one-to-three-month lags of an enhanced measure of the payout yield\footnote{Data in zipped csv, for paper \cite{Goyaletal-2024}, available here:  \url{https://sites.google.com/view/agoyal145}.} and one-to-three-month lags of the realized squared monthly volatilities are included. At each timestep, the agent can observe the 18 features including the current month and previous 11 months, as well as the logarithmic returns of the past 12 months, before making a decision. Using these data, experiments were conducted to compare three models: (i) Q-learning with a discrete state space and a discrete action space, both without leverage, $\omega \in \{0.0, 0.1,\ldots,1.0\}$, and with 50\% leverage available, $\omega \in \{0.0, 0.15,\ldots,1.5\}$; (ii) DDPG with TiDE with a continuous action space, both without leverage, $\omega \in [0.0,1.0]$, and with 50\% leverage available, $\omega \in [0.0,1.5]$; and a (iii) Buy-and-hold strategy with constant action $\omega=1.0$. We introduce buy-and-hold as a comparison baseline to determine whether active investment approaches can outperform a passive investment policy. We also introduce the opportunity for Q-learning and DDPG-TiDE strategies to use leverage (i.e., to select values $w>1$) to see if they can learn strategies that profit from borrowing to invest.

%%%%%%%%%%%%%%%%%%%%%%%%
% TAB: DATA DESCRIPTION
%%%%%%%%%%%%%%%%%%%%%%%%
\begin{table}[tb]
    \centering
    \captionsetup{justification=centering} % Ensure caption is centered
\caption{Macroeconomic and financial predictors for market environment indicators.}
    \label{tab:predictors}
    \resizebox{0.3\linewidth}{!}{%
    \begin{tabular}{ll}
        \toprule
         Predictor & Description\\
         \midrule
         dp & dividend-price ratio \\
         ep & earnings-price ratio \\
         bm & book-to-market ratio \\
         ntis & net equity expansion \\
         tbl & Treasury-bill rate \\
         tms & term spread \\
         dfy & default spread \\
         infl & inflation \\
         corpr & high-quality corporate bond rate \\
         ltr & long-term rate of return \\
         svar & stock variance \\
         \bottomrule
    \end{tabular}
    } %end resizebox
\end{table}
%%%%%%%%%%%%%%%%%%%%%%

The Q-learning and DDPG algorithms share a common dataset and trading environment module, although they differ in their state and action representations. Q-learning utilizes a discrete state and action space, achieved through K-means clustering and fixed weight increments, while DDPG operates in continuous spaces, employing a multi-step neural network time-series training approach. Furthermore, DDPG incorporates Ornstein–Uhlenbeck (OU) noise to facilitate exploration, as its policy network is inherently deterministic. A multi-step replay buffer, which stores transitions ($s,a,r,s'$), stabilizes training by breaking correlations in consecutive data and allowing mini-batch updates. The TiDE architecture in DDPG processes the time-series state before the agent encounters data: an encoder captures temporal features, while actor and critic networks propose actions and evaluate Q-values, respectively. Grid-search is used to explore the candidate range of hyperparameters. 

The performance of each algorithm, optimized with its respective hyperparameters, is evaluated once on the test set and quantified using three performance metrics (see Table~\ref{tab:metrics}): (i) Logarithmic utility maximizes the utility function when the reward is set as log-return by continuously adding up the rewards, that is, optimizing the final wealth under the Kelly criterion; (ii) Portfolio value measures the cumulative returns over a given interval, with $W_0=1$ as the initial investment and $W_t$ as the total portfolio value at time $t$; and (iii) Sharpe ratio is defined as the average excess return of a portfolio over the standard deviation of the excess portfolio return, and is calculated over a 12-month rolling window.

%%%%%%%%%%%%%%%%%%%%%%%%
% TAB: PERFORMANCE METRICS
%%%%%%%%%%%%%%%%%%%%%%%%
\begin{table}[tb]
    \caption{Performance metrics.}
    \label{tab:metrics}
    \centering
    \renewcommand{\arraystretch}{1.70}
    \resizebox{0.70\linewidth}{!}{%
    \begin{tabular}{ll}
        \toprule
         Metric & Calculation\\
         \midrule
         Logarithmic Utility (cumulative rewards) & $\textrm{LU} = \ln \left(W_{t+1}\right)=\sum_{t=1}^{T} \ln \left(1+r_{t}\right)=\sum_{t=1}^{T} reward_t$\\
         Portfolio Value (cumulative returns) & $\textrm{PV}=\frac{W_{t+1}}{W_{0}}$\\
         Sharpe Ratio (12-month rolling window) & $\textrm{SR}=\frac{\mathbb{E}_{t}\left[\rho_{t}-\rho_{f}\right]}{\sqrt{\operatorname{var}_{t}\left[\rho_{t}-\rho_{f}\right]}}$\\
         \bottomrule
    \end{tabular}
    } %end resizebox
\end{table}
%%%%%%%%%%%%%%%%%%%%%%

%%%%%%%%%%%%%%%%%%%%%%%%%%%%%%%%%%%%%%%%%%%%%
% FIG: Results
%%%%%%%%%%%%%%%%%%%%%%%%%%%%%%%%%%%%%%%%%%%%%
\begin{figure}[t]
\centering
\includegraphics[width=.7\textwidth]{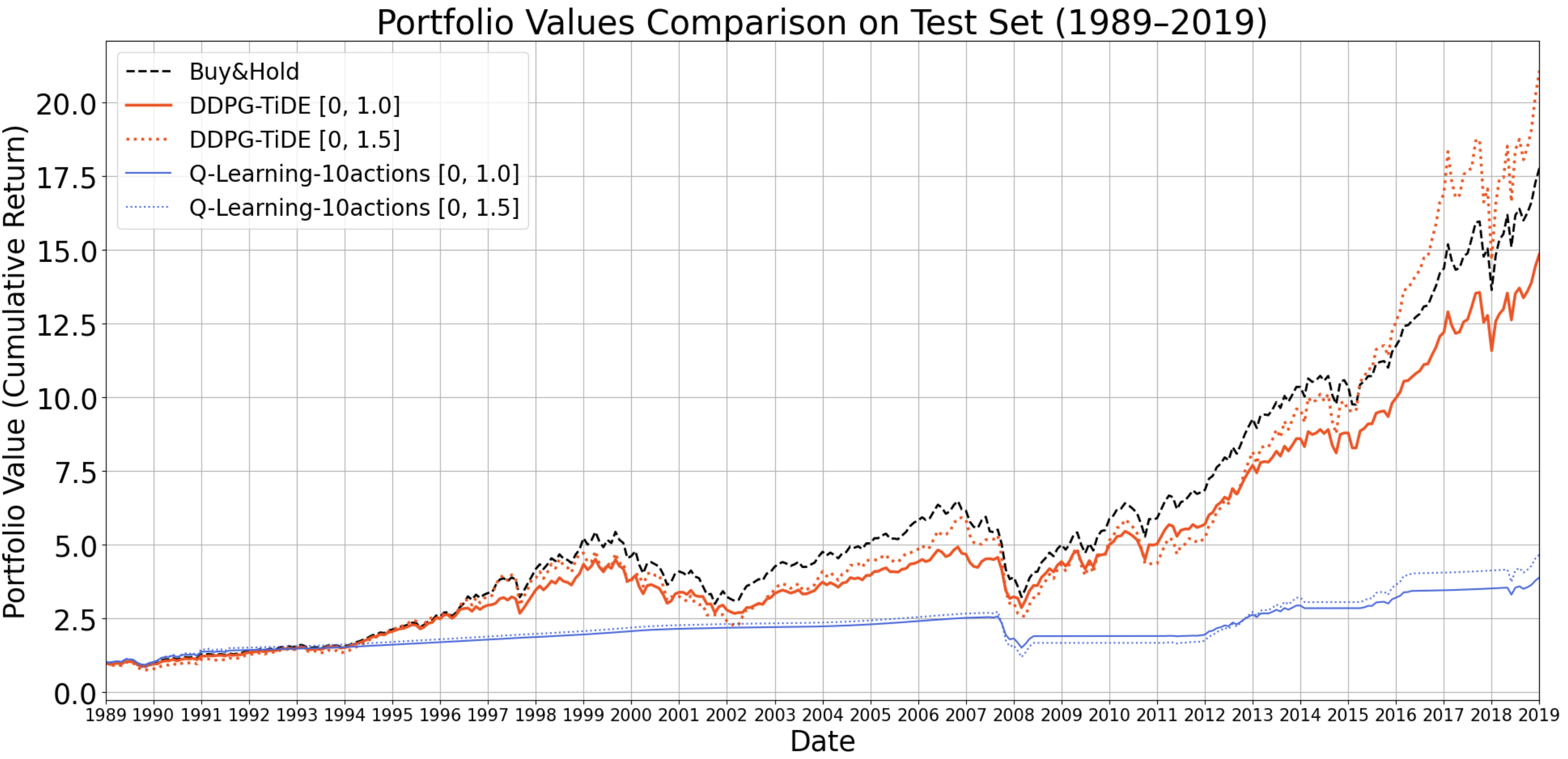}
\caption{Portfolio values comparison on test set.}
\label{fig:portfolio-value-test}
\end{figure}

\begin{figure}[t]
\centering
\includegraphics[width=.70\textwidth]{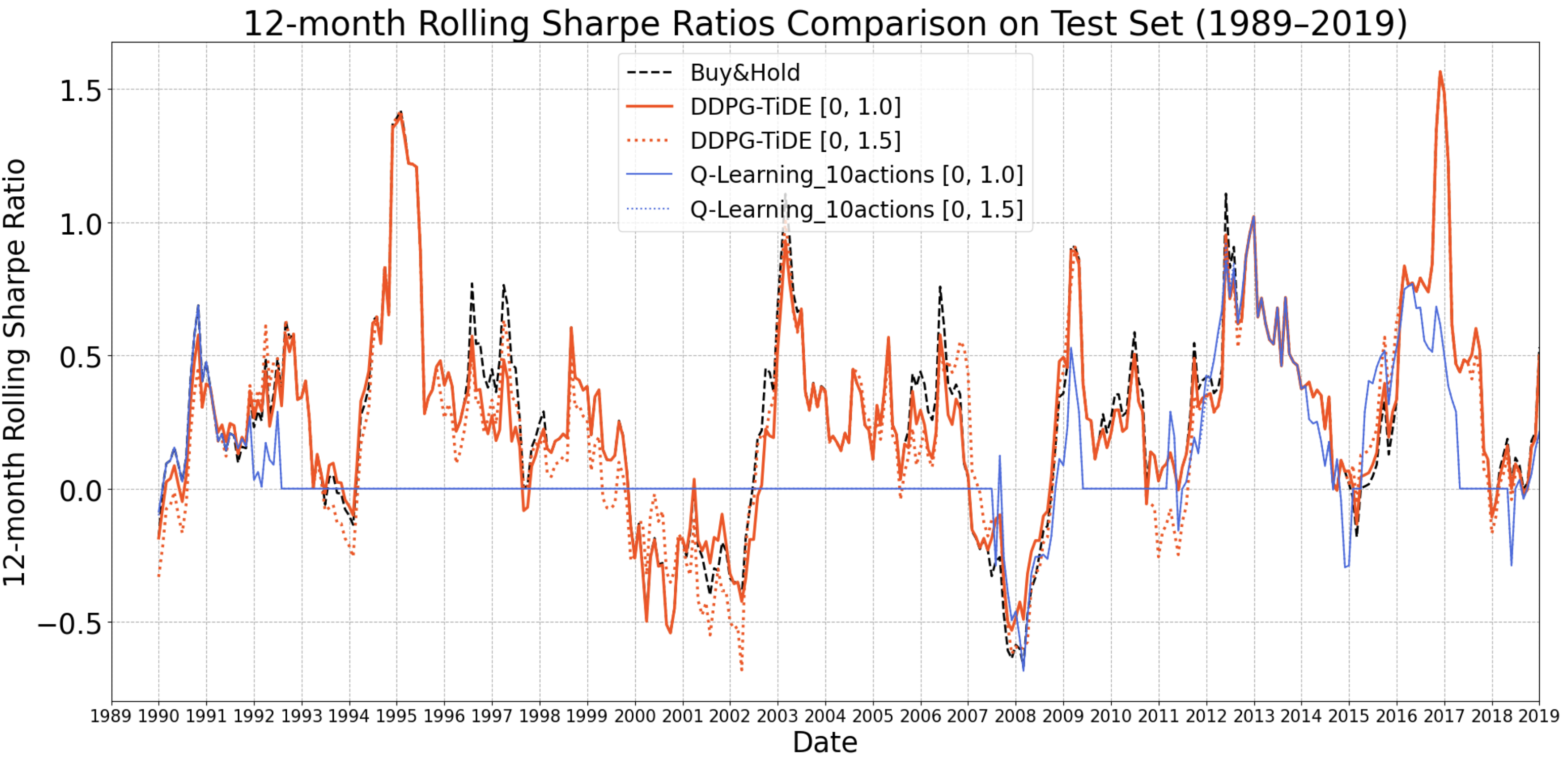}
\caption{12-month rolling window Sharpe ratios comparison on test set.}
\label{fig:sharpe-test}
\end{figure}

\begin{figure}[t]
\centering
\includegraphics[width=.70\textwidth]{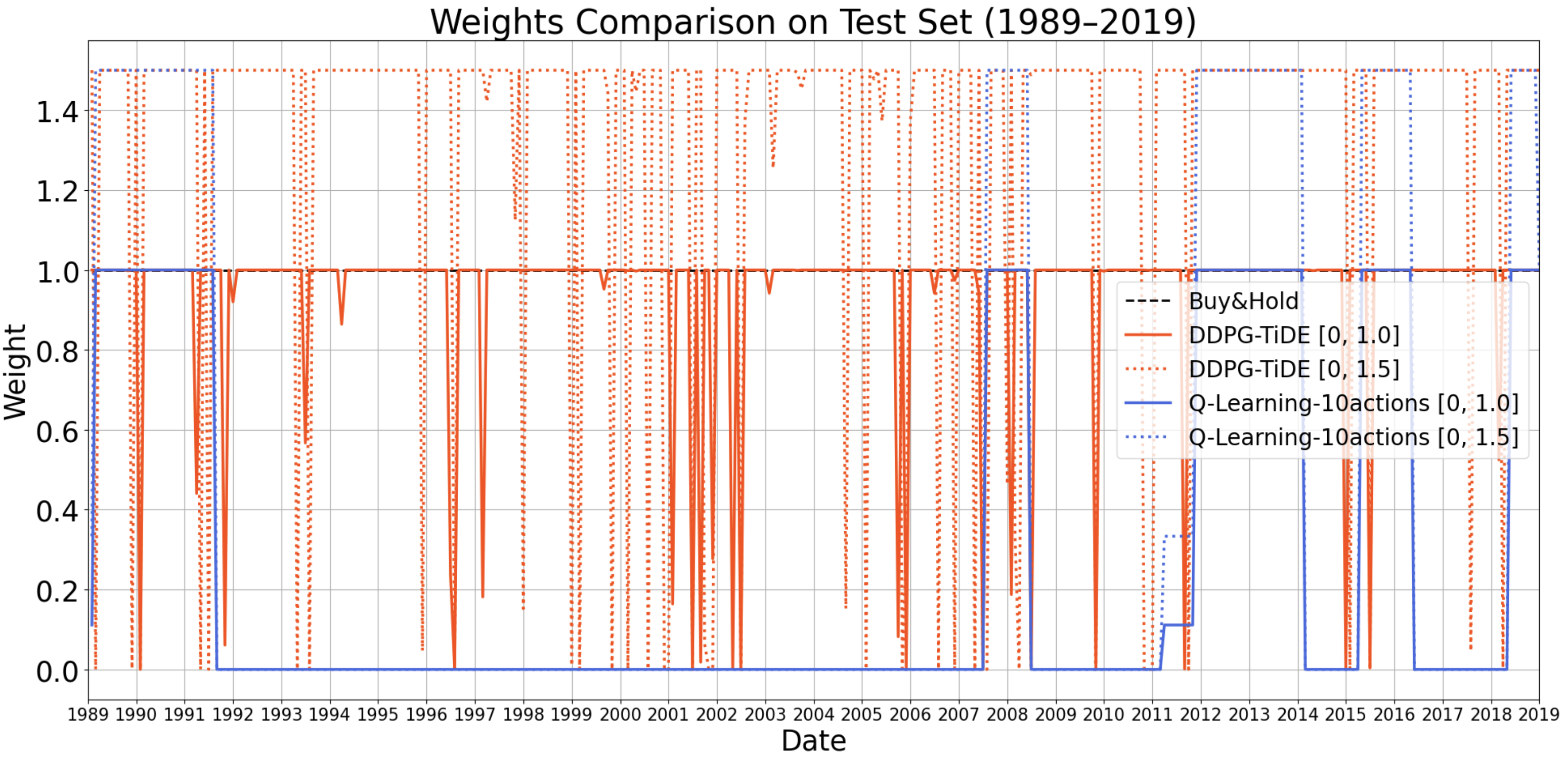}
\caption{Weights comparison on test set.}
\label{fig:weights-test}
\end{figure}
%%%%%%%%%%%%%%%%%%%%%%%%%%%%%%%%%%%%%%%%%%%%%

\section{Results and Discussion}\label{sec:results}
\noindent
In this section, we present results during the test period (1989--2019). As expected, we observe that the cumulative rewards (or logarithmic utility) and the total portfolio value (or total wealth, or cumulative returns) for each strategy follow identical trends, as the reward is simply a logarithmic transformation of wealth. Therefore, to avoid duplication, we do not present or discuss cumulative rewards. 

In Fig.~\ref{fig:portfolio-value-test}, we see that Q-learning strategies tend to generate a steady, shallow increase in portfolio value over long periods of time, particularly before 2008. The final portfolio value, or wealth, for Q-learning, both without leverage (3.88) and with 50\% leverage available (4.65), is much lower than the buy-and-hold strategy (17.77). In contrast, the portfolio values of the DDPG-TiDE strategies closely follow the value of buy-and-hold. The final portfolio value for DDPG-TiDE without leverage (14.85) is slightly below buy-and-hold, and the final portfolio value for DDPG-TiDE with 50\% leverage available (21.06) is slightly above buy-and-hold. These trends indicate that Q-learning strategies largely follow a risk-averse strategy, by predominately investing in the risk-free asset; while DDPG-TiDE strategies largely follow a risky strategy, by predominately investing in the market index.

Fig.~\ref{fig:sharpe-test} plots the Sharpe ratio of each strategy over a rolling 12-month window. For Q-learning strategies, which tend to invest in the risk-free asset, Sharpe ratios are much lower on average and are zero for long periods (e.g., 1992--2007). In contrast, Sharpe ratios for DDPG-TiDE strategies closely track buy-and-hold, with average Sharpe ratio of 1.13 when no leverage is available, and average Sharpe ratio of 0.99 when leverage is available. Both of these values are slightly higher than the average Sharpe ratio for buy-and-hold (0.95), indicating that active DDPG-TiDE strategies generate slightly better risk-adjusted returns than a passive buy-and-hold index tracker. Furthermore, we note that while the use of leverage allows DDPG-TiDE to accumulate the highest returns (see Fig.~\ref{fig:portfolio-value-test}), the increased risk of using leverage results in a slightly lower average Sharpe ratio.

We can concretely understand the performance of each strategy by observing the action taken by each strategy at each timestep. Fig.~\ref{fig:weights-test} shows the portfolio weight assigned to the risky asset each month, such that when $w=1$ all investments are held in the market index and when $w=0$ all investments are held in the risk-free asset. For values $w>1$, the strategy leverages capital by borrowing to invest more than current total wealth; for example, a value $w=1.5$ would imply 50\% leverage such that 1.5 times current wealth is invested in the risky asset. We see that the Q-learning strategy invests fully in the risk-free asset for long periods (e.g., 1992--2007), which coincides with the period of zero Sharpe ratio shown in Fig.~\ref{fig:sharpe-test}, and the smoothly increasing returns with shallow gradient shown in Fig.~\ref{fig:portfolio-value-test}.  In contrast, during this same period, DDPG-TiDE often invests fully in the risky asset, leading to a more erratic Sharpe ratio that correlates with buy-and-hold (Fig.~\ref{fig:sharpe-test}). Overall, this explains why the wealth of DDPG-TiDE largely tracks buy-and-hold, and also explains how DDPG-TiDE uses leverage (often taking values $w=1.5$) to take advantage of the bull market period post 2008 to outstrip the returns of buy-and-hold (see Fig.~\ref{fig:portfolio-value-test}).

\section{Conclusions}\label{sec:conclusion}
\noindent
We have explored the use of RL as a model-free approach to optimize portfolio allocation between a risk-free asset and a market index. 
The novel contribution of this work is the introduction of a TiDE-based architecture into the DDPG RL framework. We showed that a continuous-action DDPG-TiDE strategy outperforms a simple discrete-action Q-learning strategy in all metrics; it is also able to generate more profits than buy-and-hold by applying leverage during bull periods; and it generates the highest average Sharpe ratio. These results suggest that the integration of TiDE within a DDPG RL framework to tackle the optimal asset allocation problem is a fruitful avenue for further exploration. 

We have deliberately used a simplistic simulation framework. To advance this line of research, future work will incorporate a wider range of RL frameworks and methodologies. In addition, practical trading costs will be taken into account, and penalty mechanisms will be introduced in both algorithm design and code implementation to ensure the practicality and generalizability of the strategy. 
% Some sentences on AI4CI to close the paper
Our long-term goal is to use a multi-agent modeling approach (e.g., \cite{duffin-cifer-latency-arb-2018}) to develop intelligent collectives of reinforcement learning agents that offer user-specific personalized investment strategies (for details, refer to the AI for Collective Intelligence research strategy \cite{ai4ci-position-2024-fullnames}).

\section*{Acknowledgements}
\noindent 
This work was supported by UKRI EPSRC Grant No. EP/Y028392/1: AI for Collective Intelligence (AI4CI).

%% References
%%
%% Following citation commands can be used in the body text:
%% Usage of \cite is as follows:
%%   \cite{key}         ==>>  [#]
%%   \cite[chap. 2]{key} ==>> [#, chap. 2]
%%

%The citation must be used in following style: \cite{article-minimal} \cite{article-full} \cite{article-crossref} \cite{whole-journal}.
%% References with BibTeX database:

\bibliography{paper-refs}

\begin{thebibliography}{29}
\expandafter\ifx\csname natexlab\endcsname\relax\def\natexlab#1{#1}\fi
\providecommand{\url}[1]{\texttt{#1}}
\providecommand{\href}[2]{#2}
\providecommand{\path}[1]{#1}
\providecommand{\DOIprefix}{doi:}
\providecommand{\ArXivprefix}{arXiv:}
\providecommand{\URLprefix}{URL: }
\providecommand{\Pubmedprefix}{pmid:}
\providecommand{\doi}[1]{\href{http://dx.doi.org/#1}{\path{#1}}}
\providecommand{\Pubmed}[1]{\href{pmid:#1}{\path{#1}}}
\providecommand{\bibinfo}[2]{#2}
\ifx\xfnm\relax \def\xfnm[#1]{\unskip,\space#1}\fi
%Type = Article
\bibitem[{Aboussalah et~al.(2021)Aboussalah, Xu and Lee}]{Aboussalah2021What}
\bibinfo{author}{Aboussalah, A.}, \bibinfo{author}{Xu, Z.}, \bibinfo{author}{Lee, C.G.}, \bibinfo{year}{2021}.
\newblock \bibinfo{title}{What is the value of the cross-sectional approach to deep reinforcement learning?}
\newblock \bibinfo{journal}{Quantitative Finance} \bibinfo{volume}{22}, \bibinfo{pages}{1--21}.
\newblock \DOIprefix\doi{10.1080/14697688.2021.2001032}.
%Type = Inproceedings
\bibitem[{Bhakar et~al.(2023)Bhakar, Deori, Gautam and KG}]{Bhakar2023MaxRet}
\bibinfo{author}{Bhakar, A.}, \bibinfo{author}{Deori, P.S.}, \bibinfo{author}{Gautam, Y.V.}, \bibinfo{author}{KG, S.}, \bibinfo{year}{2023}.
\newblock \bibinfo{title}{Maximizing returns with reinforcement learning: A paradigm shift in stock market portfolio management}, in: \bibinfo{booktitle}{IEEE Region 10 Conference (TENCON)}, pp. \bibinfo{pages}{393--398}.
\newblock \DOIprefix\doi{10.1109/TENCON58879.2023.10322511}.
%Type = Article
\bibitem[{Britten-Jones(2002)}]{Britten2002MV}
\bibinfo{author}{Britten-Jones, M.}, \bibinfo{year}{2002}.
\newblock \bibinfo{title}{The sampling error in estimates of mean‐variance efficient portfolio weights}.
\newblock \bibinfo{journal}{Journal of Finance} \bibinfo{volume}{54}, \bibinfo{pages}{655--671}.
\newblock \DOIprefix\doi{10.1111/0022-1082.00120}.
%Type = Article
\bibitem[{Bullock et~al.(2024)Bullock, Ajmeri, Batty, Black, Cartlidge, Challen, Chen, Chen, Condell, Danon, Dennett, Heppenstall, Marshall, Morgan, O'Kane, Smith, Smith and Williams}]{ai4ci-position-2024-fullnames}
\bibinfo{author}{Bullock, S.}, \bibinfo{author}{Ajmeri, N.}, \bibinfo{author}{Batty, M.}, \bibinfo{author}{Black, M.}, \bibinfo{author}{Cartlidge, J.}, \bibinfo{author}{Challen, R.}, \bibinfo{author}{Chen, C.}, \bibinfo{author}{Chen, J.}, \bibinfo{author}{Condell, J.}, \bibinfo{author}{Danon, L.}, \bibinfo{author}{Dennett, A.}, \bibinfo{author}{Heppenstall, A.}, \bibinfo{author}{Marshall, P.}, \bibinfo{author}{Morgan, P.}, \bibinfo{author}{O'Kane, A.}, \bibinfo{author}{Smith, L.G.E.}, \bibinfo{author}{Smith, T.}, \bibinfo{author}{Williams, H.T.P.}, \bibinfo{year}{2024}.
\newblock \bibinfo{title}{Artificial intelligence for collective intelligence: A national-scale research strategy}.
\newblock \bibinfo{journal}{Knowledge Engineering Review} \bibinfo{volume}{39, e10}.
\newblock \DOIprefix\doi{10.1017/S0269888924000110}.
%Type = Article
\bibitem[{Cong et~al.(2020)Cong, Tang, Wang and Zhang}]{Cong2021Alphaportfolio}
\bibinfo{author}{Cong, L.W.}, \bibinfo{author}{Tang, K.}, \bibinfo{author}{Wang, J.}, \bibinfo{author}{Zhang, Y.}, \bibinfo{year}{2020}.
\newblock \bibinfo{title}{{AlphaPortfolio}: Direct construction through deep reinforcement learning and interpretable ai}.
\newblock \bibinfo{journal}{SSRN} \DOIprefix\doi{10.2139/ssrn.3554486}.
%Type = Article
\bibitem[{Das et~al.(2023)Das, Kong, Leach, Mathur, Sen and Yu}]{Das2023LongtermFW}
\bibinfo{author}{Das, A.}, \bibinfo{author}{Kong, W.}, \bibinfo{author}{Leach, A.B.}, \bibinfo{author}{Mathur, S.}, \bibinfo{author}{Sen, R.}, \bibinfo{author}{Yu, R.}, \bibinfo{year}{2023}.
\newblock \bibinfo{title}{Long-term forecasting with tide: Time-series dense encoder}.
\newblock \bibinfo{journal}{ArXiv:2304.08424} \DOIprefix\doi{10.48550/arXiv.2304.08424}.
%Type = Inproceedings
\bibitem[{Duffin and Cartlidge(2018)}]{duffin-cifer-latency-arb-2018}
\bibinfo{author}{Duffin, M.}, \bibinfo{author}{Cartlidge, J.}, \bibinfo{year}{2018}.
\newblock \bibinfo{title}{Agent-based model exploration of latency arbitrage in fragmented financial markets}, in: \bibinfo{booktitle}{IEEE Symposium on Computational Intelligence for Financial Engineering and Economics (CIFEr)}, pp. \bibinfo{pages}{2312--2320}.
\newblock \DOIprefix\doi{10.1109/SSCI.2018.8628638}.
%Type = Article
\bibitem[{Goyal et~al.(2024)Goyal, Welch and Zafirov}]{Goyaletal-2024}
\bibinfo{author}{Goyal, A.}, \bibinfo{author}{Welch, I.}, \bibinfo{author}{Zafirov, A.}, \bibinfo{year}{2024}.
\newblock \bibinfo{title}{A comprehensive 2022 look at the empirical performance of equity premium prediction}.
\newblock \bibinfo{journal}{Review of Financial Studies} \bibinfo{volume}{37(11)}, \bibinfo{pages}{3490--3557}.
\newblock \DOIprefix\doi{10.1093/rfs/hhae044}.
%Type = Article
\bibitem[{Hambly et~al.(2023)Hambly, Xu and Yang}]{Hambly2023Recent}
\bibinfo{author}{Hambly, B.}, \bibinfo{author}{Xu, R.}, \bibinfo{author}{Yang, H.}, \bibinfo{year}{2023}.
\newblock \bibinfo{title}{Recent advances in reinforcement learning in finance}.
\newblock \bibinfo{journal}{Mathematical Finance} \bibinfo{volume}{33}, \bibinfo{pages}{437--503}.
\newblock \DOIprefix\doi{https://doi.org/10.1111/mafi.12382}.
%Type = Article
\bibitem[{Jiang et~al.(2022)Jiang, Saunders and Weng}]{Jiang2022Kelly}
\bibinfo{author}{Jiang, R.}, \bibinfo{author}{Saunders, D.}, \bibinfo{author}{Weng, C.}, \bibinfo{year}{2022}.
\newblock \bibinfo{title}{The reinforcement learning kelly strategy}.
\newblock \bibinfo{journal}{Quantitative Finance} \bibinfo{volume}{22(8)}, \bibinfo{pages}{1445--1464}.
\newblock \DOIprefix\doi{10.1080/14697688.2022.204935}.
%Type = Article
\bibitem[{Jiang et~al.(2017)Jiang, Xu and Liang}]{Jiang2017ADR}
\bibinfo{author}{Jiang, Z.}, \bibinfo{author}{Xu, D.}, \bibinfo{author}{Liang, J.}, \bibinfo{year}{2017}.
\newblock \bibinfo{title}{A deep reinforcement learning framework for the financial portfolio management problem}.
\newblock \bibinfo{journal}{ArXiv:1706.10059} \DOIprefix\doi{10.48550/arXiv.1706.10059}.
%Type = Article
\bibitem[{Kan and Zhou(2007)}]{KanZhou2007Optimal}
\bibinfo{author}{Kan, R.}, \bibinfo{author}{Zhou, G.}, \bibinfo{year}{2007}.
\newblock \bibinfo{title}{Optimal portfolio choice with parameter uncertainty}.
\newblock \bibinfo{journal}{Journal of Financial and Quantitative Analysis} \bibinfo{volume}{42}, \bibinfo{pages}{621--656}.
\newblock \DOIprefix\doi{10.1017/S0022109000004129}.
%Type = Article
\bibitem[{Liang et~al.(2018)Liang, Chen, Zhu, Jiang and Li}]{Liang2018AdversarialDR}
\bibinfo{author}{Liang, Z.}, \bibinfo{author}{Chen, H.}, \bibinfo{author}{Zhu, J.}, \bibinfo{author}{Jiang, K.}, \bibinfo{author}{Li, Y.}, \bibinfo{year}{2018}.
\newblock \bibinfo{title}{Adversarial deep reinforcement learning in portfolio management}.
\newblock \bibinfo{journal}{ArXiv:1808.09940v3} \DOIprefix\doi{10.48550/arXiv.1808.09940}.
%Type = Article
\bibitem[{Lillicrap et~al.(2015)Lillicrap, Hunt, Pritzel, Heess, Erez, Tassa, Silver and Wierstra}]{Lillicrap2015ContinuousCW}
\bibinfo{author}{Lillicrap, T.P.}, \bibinfo{author}{Hunt, J.J.}, \bibinfo{author}{Pritzel, A.}, \bibinfo{author}{Heess, N.M.O.}, \bibinfo{author}{Erez, T.}, \bibinfo{author}{Tassa, Y.}, \bibinfo{author}{Silver, D.}, \bibinfo{author}{Wierstra, D.}, \bibinfo{year}{2015}.
\newblock \bibinfo{title}{Continuous control with deep reinforcement learning}.
\newblock \bibinfo{journal}{ArXiv:1509.02971} \DOIprefix\doi{10.48550/arXiv.1509.02971}.
%Type = Incollection
\bibitem[{MacLean et~al.(2010)MacLean, Thorp, Zhao and Ziemba}]{Maclean2010Medium}
\bibinfo{author}{MacLean, L.C.}, \bibinfo{author}{Thorp, E.O.}, \bibinfo{author}{Zhao, Y.}, \bibinfo{author}{Ziemba, W.T.}, \bibinfo{year}{2010}.
\newblock \bibinfo{title}{Medium term simulations of the full {Kelly} and fractional {Kelly} investment strategies}, in: \bibinfo{booktitle}{The Kelly Capital Growth Investment Criterion}. \bibinfo{publisher}{World Scientific}, pp. \bibinfo{pages}{543--561}.
\newblock \DOIprefix\doi{10.1142/9789814293501_0038}.
%Type = Article
\bibitem[{MacLean et~al.(2011)MacLean, Thorp and Ziemba}]{Maclean2011Longterm}
\bibinfo{author}{MacLean, L.C.}, \bibinfo{author}{Thorp, E.O.}, \bibinfo{author}{Ziemba, W.T.}, \bibinfo{year}{2011}.
\newblock \bibinfo{title}{Long-term capital growth: the good and bad properties of the {Kelly} and fractional {Kelly} capital growth criteria}.
\newblock \bibinfo{journal}{Quantitative Finance} \bibinfo{volume}{10}, \bibinfo{pages}{681--687}.
\newblock \DOIprefix\doi{10.1080/14697688.2010.506108}.
%Type = Article
\bibitem[{Markowitz(1952)}]{Markowitz1952}
\bibinfo{author}{Markowitz, H.}, \bibinfo{year}{1952}.
\newblock \bibinfo{title}{{Portfolio Selection}}.
\newblock \bibinfo{journal}{The Journal of Finance} \bibinfo{volume}{7}, \bibinfo{pages}{77--91}.
\newblock \DOIprefix\doi{10.1111/j.1540-6261.1952.tb01525.x}.
%Type = Article
\bibitem[{Merton(1969)}]{Merton1969Lifetime}
\bibinfo{author}{Merton, R.C.}, \bibinfo{year}{1969}.
\newblock \bibinfo{title}{Lifetime portfolio selection under uncertainty: The continuous-time case}.
\newblock \bibinfo{journal}{The Review of Economics and Statistics} \bibinfo{volume}{51}, \bibinfo{pages}{247--257}.
\newblock \DOIprefix\doi{10.2307/1926560}.
%Type = Incollection
\bibitem[{Merton(1971)}]{Merton1971Optimum}
\bibinfo{author}{Merton, R.C.}, \bibinfo{year}{1971}.
\newblock \bibinfo{title}{Optimum consumption and portfolio rules in a continuous-time model}, in: \bibinfo{booktitle}{Stochastic optimization models in finance}. \bibinfo{publisher}{Elsevier}, pp. \bibinfo{pages}{621--661}.
\newblock \DOIprefix\doi{10.1016/0022-0531(71)90038-X}.
%Type = Article
\bibitem[{Park et~al.(2020)Park, Sim and Choi}]{Park2020AIFP}
\bibinfo{author}{Park, H.}, \bibinfo{author}{Sim, M.K.}, \bibinfo{author}{Choi, D.G.}, \bibinfo{year}{2020}.
\newblock \bibinfo{title}{An intelligent financial portfolio trading strategy using deep {Q}-learning}.
\newblock \bibinfo{journal}{Expert Systems with Applications} \bibinfo{volume}{158}, \bibinfo{pages}{113573}.
\newblock \DOIprefix\doi{10.1016/j.eswa.2020.113573}.
%Type = Article
\bibitem[{Pendharkar and Cusatis(2018)}]{Pendharkar2018Trading}
\bibinfo{author}{Pendharkar, P.C.}, \bibinfo{author}{Cusatis, P.}, \bibinfo{year}{2018}.
\newblock \bibinfo{title}{Trading financial indices with reinforcement learning agents}.
\newblock \bibinfo{journal}{Expert Systems with Applications} \bibinfo{volume}{103}, \bibinfo{pages}{1--13}.
\newblock \DOIprefix\doi{10.1016/j.eswa.2018.02.032}.
%Type = Article
\bibitem[{Pham et~al.(2021)Pham, Luu and Tran}]{Pham2021MultiagentRL}
\bibinfo{author}{Pham, U.H.}, \bibinfo{author}{Luu, Q.C.}, \bibinfo{author}{Tran, H.D.}, \bibinfo{year}{2021}.
\newblock \bibinfo{title}{Multi-agent reinforcement learning approach for hedging portfolio problem}.
\newblock \bibinfo{journal}{Soft {C}omputing} \bibinfo{volume}{25}, \bibinfo{pages}{7877--7885}.
\newblock \DOIprefix\doi{10.1007/s00500-021-05801-6}.
%Type = Article
\bibitem[{Pinelis and Ruppert(2022)}]{Pinelis2022Machine}
\bibinfo{author}{Pinelis, M.}, \bibinfo{author}{Ruppert, D.}, \bibinfo{year}{2022}.
\newblock \bibinfo{title}{Machine learning portfolio allocation}.
\newblock \bibinfo{journal}{The Journal of Finance and Data Science} \bibinfo{volume}{8}, \bibinfo{pages}{35--54}.
\newblock \DOIprefix\doi{10.1016/j.jfds.2021.12.001}.
%Type = Article
\bibitem[{Simon et~al.(2022)Simon, Weibels and Zimmermann}]{Simon2022DeepPP}
\bibinfo{author}{Simon, F.}, \bibinfo{author}{Weibels, S.}, \bibinfo{author}{Zimmermann, T.}, \bibinfo{year}{2022}.
\newblock \bibinfo{title}{Deep parametric portfolio policies}.
\newblock \bibinfo{journal}{SSRN} \DOIprefix\doi{10.2139/ssrn.4150292}.
%Type = Article
\bibitem[{Watkins and Dayan(1992)}]{Watkins1992Q}
\bibinfo{author}{Watkins, C.J.}, \bibinfo{author}{Dayan, P.}, \bibinfo{year}{1992}.
\newblock \bibinfo{title}{Q-learning}.
\newblock \bibinfo{journal}{Machine learning} \bibinfo{volume}{8}, \bibinfo{pages}{279--292}.
\newblock \DOIprefix\doi{10.1007/BF00992698}.
%Type = Article
\bibitem[{Wei et~al.(2019)Wei, Wang, Mangu and Decker}]{Wei2019ModelbasedRL}
\bibinfo{author}{Wei, H.}, \bibinfo{author}{Wang, Y.}, \bibinfo{author}{Mangu, L.}, \bibinfo{author}{Decker, K.S.}, \bibinfo{year}{2019}.
\newblock \bibinfo{title}{Model-based reinforcement learning for predictions and control for limit order books}.
\newblock \bibinfo{journal}{ArXiv:1910.03743} \DOIprefix\doi{10.48550/arXiv.1910.03743}.
%Type = Article
\bibitem[{Yu et~al.(2019)Yu, Lee, Kulyatin, Shi and Dasgupta}]{Yu2019ModelbasedDR}
\bibinfo{author}{Yu, P.}, \bibinfo{author}{Lee, J.S.}, \bibinfo{author}{Kulyatin, I.}, \bibinfo{author}{Shi, Z.}, \bibinfo{author}{Dasgupta, S.}, \bibinfo{year}{2019}.
\newblock \bibinfo{title}{Model-based deep reinforcement learning for dynamic portfolio optimization}.
\newblock \bibinfo{journal}{ArXiv:1901.08740} \DOIprefix\doi{10.48550/arXiv.1901.08740}.
%Type = Misc
\bibitem[{Zhang et~al.(2021)Zhang, Zhang, Cucuringu and Zohren}]{Zhang2021UniversalE2E}
\bibinfo{author}{Zhang, C.}, \bibinfo{author}{Zhang, Z.}, \bibinfo{author}{Cucuringu, M.}, \bibinfo{author}{Zohren, S.}, \bibinfo{year}{2021}.
\newblock \bibinfo{title}{A universal end-to-end approach to portfolio optimization via deep learning}.
\newblock \DOIprefix\doi{10.48550/arXiv.2111.09170}.
%Type = Article
\bibitem[{Zhao et~al.(2025)Zhao, Zhang, Qin and Yang}]{Zhao2025QuantFactor}
\bibinfo{author}{Zhao, J.}, \bibinfo{author}{Zhang, C.}, \bibinfo{author}{Qin, M.}, \bibinfo{author}{Yang, P.}, \bibinfo{year}{2025}.
\newblock \bibinfo{title}{Quantfactor reinforce: Mining steady formulaic alpha factors with variance-bounded reinforce}.
\newblock \bibinfo{journal}{IEEE Transactions on Signal Processing} , \bibinfo{pages}{1--16}\DOIprefix\doi{10.1109/tsp.2025.3576781}.

\end{thebibliography}
\bibliographystyle{elsarticle-harv}

\end{document}